\documentclass[twocolumn,amssymb, nobibnotes, aps, prb,groupedaddress]{revtex4-1}
\usepackage{graphicx}

\begin{document}
\begin{abstract}
Fluorine adatoms on graphene induce local changes in electronic and magnetic properties, and subtle correlation effects. We investigate the GGA and GGA+U approaches as possible solutions to describe the magnetic moment and electronic band structure of graphene sheets with fluorine adatoms, and compare to experiments. We show that, due to a lack of strong electronic correlations, GGA fails to reproduce the measured magnetic moment in this structure. In particular, the GGA incorrectly predicts a nonmagnetic ground state with a zero band gap. On the other hand, GGA+U is a computationally efficient tool which provides physically reasonable properties. Using Hubbard U and exchange J parameters of 5 eV and 0.1 eV provides a magnetic moment and optical gap in agreement with experiments. Our results imply that the magnetic moment observed in the experiment is injected by fluorine in carbon p$_z$ orbitals throughout the graphene sheet. The spin-orbit coupling (SOC) has almost no influence (ca. 2\%) on the magnetism. No Rashba effect is detected and the magnetic moment induced by fluorine strongly dominates the electronic properties. Our findings explain the anisotropic magnetic behavior observed experimentally.
\end{abstract}

\title{Localization of electrons and magnetization in fluoro-graphene: A DFT+U study}

\author{F. Marsusi}%
\email{marsusi@aut.ac.ir}
\affiliation{$^1$Department of Physics, Amirkabir University of Technology, P.O. Box 15875-4413, Tehran, Iran
}
\author{M. J. Verstraete}
\email{matthieu.verstraete@ulg.ac.be}
\affiliation{$^2$nanomat/QMAT/CESAM and European Theoretical Spectroscopy Facility,\\
                 Universit\'{e} de Li\`{e}ge, B-4000 Sart Tilman, Belgium}
\date{August 10, 2017}%
\maketitle
\section{Introduction}
Fluorinated graphene is a potential candidate for logic circuits and nanoelectronic devices due to the integration of graphene's exceptional properties with localized and quantum fluorine defects\cite{Ho, Cheng, Han}. While pristine graphene is a diamagnetic semiconductor with zero band gap, many experiments have confirmed that fluorination induces local magnetic moments on graphene and opens the band gap \cite{Nair_2012, Hong, Zhao}. Despite the large number of studies which consider the electronic properties of fluorinated graphene\cite{Irmer, Kim, Sahin, Santos}, the influence of fluorine bonding on the magnetization, specifically at low fluorine content, remains theoretically an open question\cite{Han}.

Treating fluorine adatoms on graphene (FG) within density functional theory (DFT) is a questionable approach and predicts a non-magnetic moment ground state. While an attempt was made to relate this non-magnetic ground state to the ionic character of the C--F bond\cite{Santos}, experimental measurements negate this prediction and confirm a spin-polarized ground state for FG, especially at low fluorine coverage\cite{Nair_2012, Hong}. According to the experimental evidence, dilute FG shows an unexpected colossal negative magnetoresistance comparable to those observed in ferromagnetic semiconductors. This suggests fluorine on graphene produces a strong magnetic moment and a quantum metal-insulator transition\cite{Hong}. 

DFT functionals often suffer from an electron delocalization error: they have a tendency to spread out the electron density artificially\cite{Mori}. This erroneous behavior of density is connected to many-electron self-interaction errors (SIE), which are not completely canceled by local exchange. SIE especially manifests itself for systems with odd number of electrons, resulting in an incorrect description of the energy of the orbitals and spin states.

Local and semi-local exchange-correlation functionals work well in weakly correlated systems\cite{Ceder, Casely}, but they provide a relatively poor description of many strongly correlated regimes\cite{Cohen}. It has been shown recently that the total magnetization for hydrogen adsorbed on graphene can drop to zero due to the SIE\cite{Casolo}. Hybrid functionals can partially correct for the SIE by including of a fraction of exact exchange adjusted to cancel the SIE\cite{Seminario}. One recent work which improved on the DFT description of the magnetic moment and optical gap of dilute FG was based on hybrid functionals\cite{Kim}. Generally, hybrid functionals produce a good description of high spin systems, but may fail for low spin complexes\cite{Scherlis, Verma}. In addition, depending on the system under study, these functionals could be computationally quite expensive.

DFT+U is an alternative technique to correct the SIE for localized atomic-like states. In this method, the exchange-correlation functional is modified with a correction functional, while the parameter U controls the magnitude of this correction. In the present study, we propose GGA+U to describe the electronic properties of FG, and to treat the strong correlations stemming from electron localization. DFT+U constructs a corrective functional, inspired by the Hubbard model, which presents a simple expression for the correlation term with a computational cost comparable to standard DFT calculations\cite{Himmetoglu}. We will show that the introduction of fluorine on graphene sheets does not lead to a perfect $sp^3$ rehybridization. To understand the hybridization character of carbon orbitals and the C–-F bonding nature, the GGA density of states has been calculated, projected on atomic orbitals (pDOS). The defect states are found to have some d-like character near the Fermi level. This finding further encourages us to use the more appropriate and accurate GGA+U method.

We confirm that, in disagreement with experiment, spin-polarized GGA calculations predict the ground state of FG to be a delocalized nonmagnetic configuration. The GGA+U reproduces the essential experimental features of FG. We show that $sp^3$ bonding between carbon and fluorine tends to localize the electrons responsible for $\pi$-bonding in pristine graphene. By adding F atoms, the electron density distribution around the central carbon bonded to F, C(F), changes and induces a magnetic moment in the FG sheet. If fluorine bonds to one of the carbons in the A-sublattice of graphene, the electrons at the top of the valence band will be spin-down electrons localized on the A-sublattice, while the bottom of the conduction band will be composed of spin-up electrons localized on the B sites. A gap emerges, in particular at the center of the first Brillouin zone (BZ) ($\Gamma$-point).
\begin{figure}
\includegraphics[width=1. \linewidth]{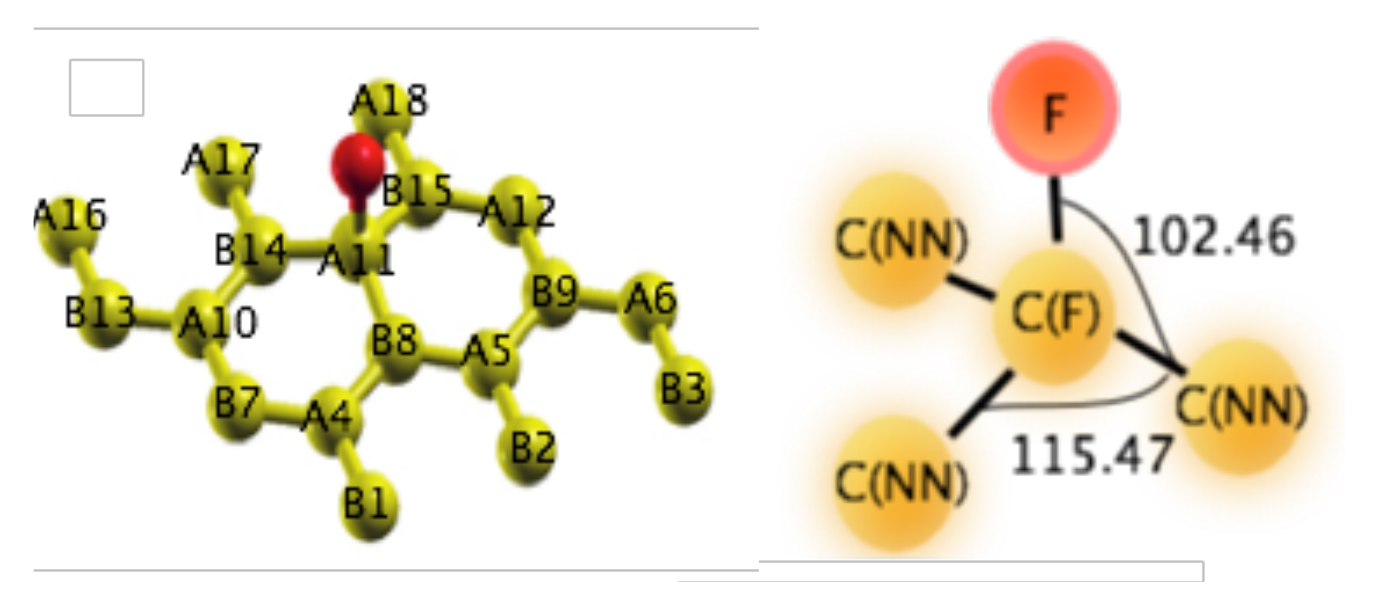}
\caption{\label{FIG1} {(Color online) Geometry of single F adatom on a 3$\times$3 supercell of graphene (FG) configuration. Fluorine is shown as a red ball. Carbons are labeled by the corresponding sublattice and number of atom. C(F) denotes the carbon atom bonded to the fluorine and its nearest neighbors are C(NN).}}
\hspace*{-1.5 cm}
\centering
\end{figure}

\begin{figure*}
\includegraphics[width=0.7 \linewidth]{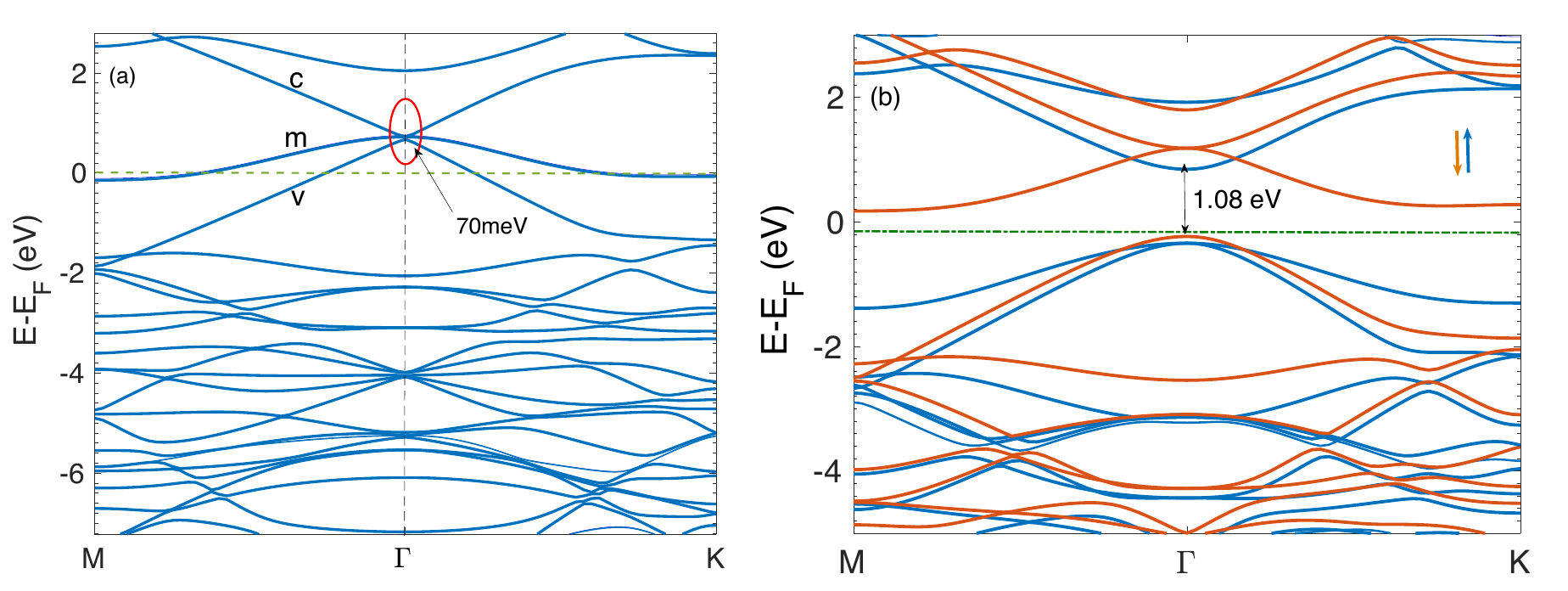}
\caption{\label{FIG2} {(Color on line) (a) PBE+PAW predicted band structure of FG. Valence (v), mid-gap (m) and conduction (c) bands are specified. Almost no spin polarization is observed. The formation of an electric dipole moment along the C--F bond and reduction of the symmetry from D$_{6h}$ in pristine graphene to C$_{3v}$ in FG lifts the degeneracy between $\pi$ and $\pi$* states and opens a 70 meV splitting in the spin majority channel at the $\Gamma$ point, but there is no overall gap between the highest occupied and the lowest unoccupied states. (b) PBE+U predicted band structure of FG. The corresponding spin up and down states are specified by blue and brown colors, respectively. A direct gap of 1.08 eV between the top of the highest occupied and the lowest unoccupied band at $\Gamma$-point is opened. The indirect gap between $\Gamma$ and M points is 0.41 eV with minority channel band edges. The Fermi energy E$_F$ set to zero. }}
\hspace*{-0.3cm}
\centering
\end{figure*}
The insulating character of the system can be explained by spin alignment of those electrons in the vicinity of Fermi level, induced by fluorine. The resulting magnetization is preferentially oriented along the symmetry axes of the structure, induced by the C--F bond and perpendicular to the graphene plane with a magnetic anisotropy energy (MAE) of 3.95 $\mu$eV/F adatom. Forcing the electrons to localize in atomic-like orbitals leads to increasing Coulomb interaction and correlation. 
We also perform GGA+U+SOC calculations and find that the SOC interaction only has a small contribution to the exchange splitting. 

\section{Computational details}
One fluorine atom is attached to a 3$\times$3 supercell of graphene at the top position of a carbon, as shown in Fig.~\ref{FIG1}, yielding an experimentally representative coverage fraction for FG. Other coverage regimes with full or very sparse fluorination are very interesting, but yield different electronic properties and will not be studied here. 
We employ a first code (ABINIT)\cite{ABINIT} to relax the structure and evaluate a first implementation of DFT+U, within PAW spheres. We compare with a second code, Quantum Espresso (QE)\cite{QE}, which has a complete implementation of the DFT+U in all space (not just in atomic spheres) -- this will prove to be important for the extended p electrons, and will be our reference. 
A (5$\times$5$\times$1) Monkhorst-Pack \textbf{k}-point mesh was used to sample the BZ \cite{Monkhorst}. The geometry is optimized within a GGA (PBE) plane-wave scheme, using norm-conserving (NC) Troullier-Martins (TM)\cite{Troullier} pseudopotentials (PP). 
The plane waves are limited with a 40 a.u. energy cutoff and all atomic positions and the unit cell are fully optimized with atomic forces below 6 meV/\AA. Electronic properties including band structure and SOC effects are investigated using the projector-augmented wave (PAW) method with QE, with a converged PAW kinetic energy cutoff of 28 a.u. Calculations are first performed using collinear spin-polarized wave functions. To calculate the strongly correlated electronic structure, we use the rotationally invariant scheme introduced by Liechtenstein et al.\cite{Lichtenstein} and then simplified by Dudarev\cite{Dudarev}. Within this formulation, the non-spherical electronic interactions are neglected. The optimized geometry obtained by PBE, when used in the PBE+U calculations shows a maximum stress of about 1.0d-4 a.u. 
Applying the commonly used approximation of $U_{eff}=U-J$ and setting $J=0$, gives a slightly smaller magnetic moment. 
We have performed $\pi$-orbital axis (POAV1) vector analysis as described in Ref.s~[\onlinecite{Haddon}] and [\onlinecite{Bai}] to evaluate the orbital nature. We have used the method introduced in Ref. [\onlinecite{Bai}], which is based on the coordinates of the conjugated central C(F) and the three attached atoms C(NN). The POAV2 analysis\cite{Haddon_2} is performed to predict the nature of the C–-F bond by calculating the degree of the p-orbital content in the $\sigma$-orbitals ($sp^n$).
Finally, we calculate spin-orbit splittings using a standard non-collinear scheme applied with two-component Pauli spinors. 

\begin{figure}
\hspace*{-0.6 cm}
\centering
\includegraphics[width=1 \linewidth]{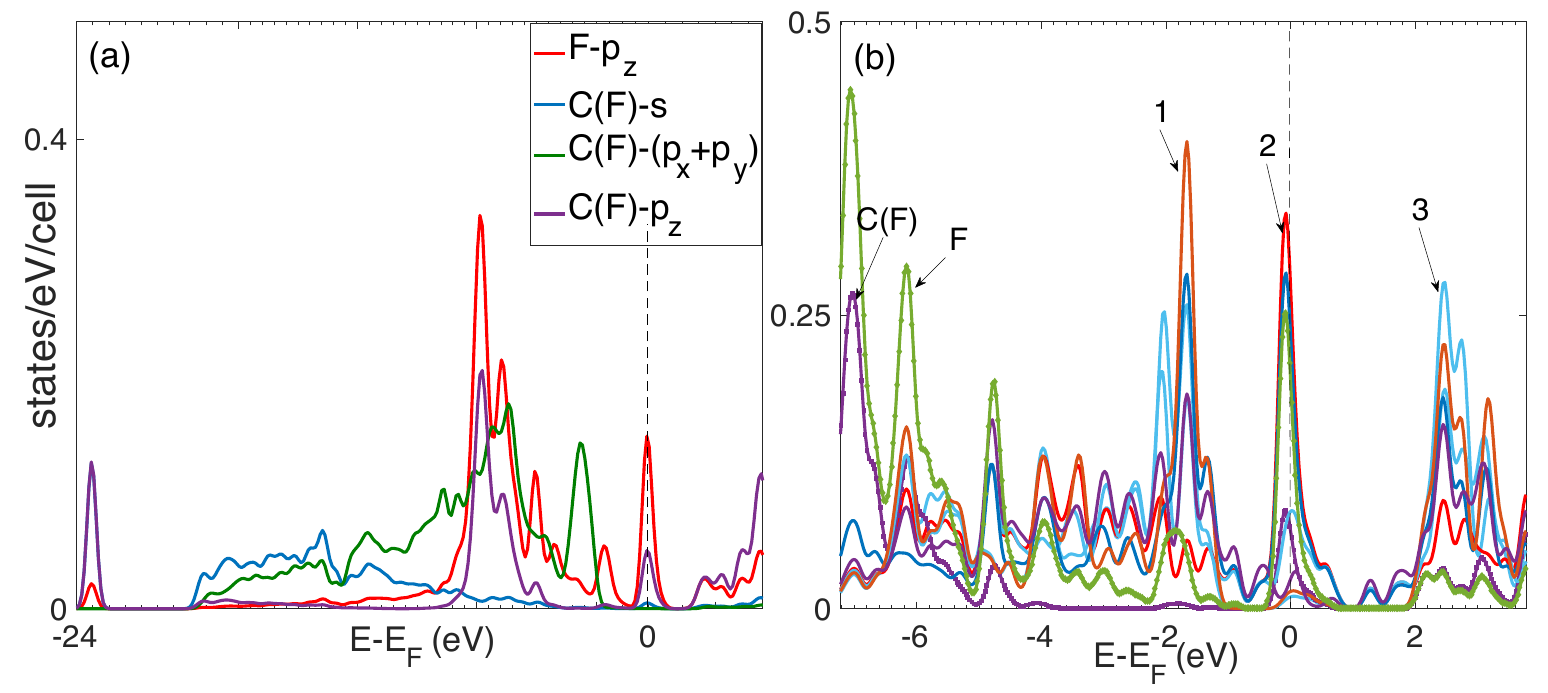}
\caption{\label{FIG3} {(Color on line) (a) GGA predicted pDOS of C(F) carbon atom of graphene along with pDOS p$_z$ of fluorine. (b) GGA predicted pDOS of p$_z$ orbital of all atoms in FG structure shows three peaks in the vicinity of the Fermi level, which are numbered by 1, 2 and 3. The middle peak, labeled by number 2, shows the contribution of p$_z$ to form mid-gap state. The Fermi energy E$_F$ is set to zero.
}}
\end{figure}

\section{Results and discussions}
\subsection{Optimized geometries and bonding hybridization}
After relaxation, as presented in Fig.~\ref{FIG1}, the single adatom remains on top of the carbon, the most favorable position for chemisorption\cite{Casolo}. C(F)-–F bond length is predicted to be about 1.55 \AA, which is larger than the typical carbon-fluorine $sp^3$ bond length (about 1.37 \AA), and is in good agreement with a previous PBE prediction\cite{Santos}. Other geometric information, including the distance between the central carbon and its nearest neighbors C(NN), is presented in Table \ref{tab1}. As a consequence of the induced  $sp^3$ hybridization between the C and F atoms, C(F) is pulled out of the plane by about 0.33 \AA. The three C(F)–-C(NN) $\sigma$--bonds of graphene resist stretching; consequently, these atoms are also dragged slightly out of the plane to reduce the bond stress. This results in a decrease in the values of the three angles ($\angle$C(NN), C(F), C(NN)) in the interior hexagonal ring around the central C(F) from 120.0$^{\circ}$ to about 115.5$^{\circ}$. 

\begin{table}[h!]
\caption{\label{tab1} PBE-optimized geometry parameters of a single F adatom on a 3$\times$3 supercell of graphene. The $z$-components and the angles between the specified atoms shown in Fig.~\ref{FIG1} are listed. The Z axis is taken normal to the initial graphene plane.The bond lengths are in \AA~and angles in degree.}
\begin{tabular}{@{}lc}
\toprule
parameter&3$\times$3\\
\hline
$z$(C(F))&0.33\\
$z$(C(NN))&0.02\\
C--F&1.55\\
C(F)-C(NN)&1.48\\
$\angle$C(NN)-C(F)-F&102.46\\
$\angle$C(NN)-C(F)-C(NN)&115.47\\
\toprule\\
\end{tabular}\\
\end{table}
$\pi$-orbital axis vector analysis predicts a deviation from $sp^2$ to $sp^{2.33}$ hybridization for the three $\sigma$-bonds pointing towards C(F) atom. Thus, by adding fluorine these three $\sigma$-bonds gain only slight $sp^3$ character. The three $\sigma$-bonds resist further pulling C(F) up and prevent the formation of three 109.5$^{\circ}$ angles between F–-C(F)–-C(NN). This results in a larger C–-F bond length and a smaller bond angle of 102.4$^{\circ}$. Analyzing the $\sigma$- and $\pi$-orbitals of atoms using the POAV2 method demonstrates the formation of $sp^{4.66}$ rehybridization in the C(F)–-F bond, which indicates the great contribution of nearby p-orbitals to this bond. Forming $sp^{4.66}$ hybridization, rather than $sp^3$, is due to the large electronegativity of fluorine on one side, and the resistance of graphene to maintain its $sp^2$ character.

\subsection{PBE Electronic structure }
The spin-polarized ground state of FG in PBE is favored by only 1 meV over the spin-unpolarized configuration (within the intrinsic error). We will see later that this stems from the PBE inability to predict correctly exchange interactions of the correlated electrons in this system. The Dirac point \textbf{K} in the primitive hexagonal BZ is folded to the $\Gamma$-point for the supercell represented in Fig. \ref{FIG1}. The PBE electronic band structure is plotted in Fig. \ref{FIG2}(a) showing a zero band gap. PBE predicts that the maximum splitting between spin up and down electrons happens for conduction (c) band (14~meV) and mid-gap (m) states (17~ meV), respectively. The spin splittings for other bands are at most 1-7 meV. According to PBE, the Fermi level is shifted down compared to the Dirac point of pristine graphene by 0.54 eV. This shift is in good agreement with the value reported in Ref.~[\onlinecite{Kim}] (0.45 eV) for larger supercells. L\"{o}wdin\cite{Lowdin} population analysis shows a relatively large charge of about 0.38 e transferred from carbon C(F) into the fluorine adatom, giving ionic character to the C–-F covalent bond. Therefore, electrons experience an electric field stemming from a dipole at the puckered out point. This feature along with the reduction of the symmetry from D$_{6h}$ in pristine graphene into C$_{3v}$ in FG lift the degeneracy between $\pi$ (at -2.13 eV, or 0.65 eV from $E_F$) and $\pi$* (at -2.06 eV or 0.72 eV from $E_F$) states and opens a small direct gap of about 74 meV from spin majority channel at the $\Gamma$ point, as illustrated in Fig. \ref{FIG2}(a). Likewise, a 91 meV gap is opened between $\pi$ (at -2.14 eV) and $\pi$* (at -2.05 eV) states from spin minority channel. There is no global gap between the highest occupied and the lowest unoccupied states.

The $sp^3$ bonding and antibonding states made of F and C(F) atoms have no contribution to the valence (v) and mid-gap bands, and are located in the intervals (-7, -5) and (3, 5) eV with respect to the Fermi level. The four orbitals of C(F) (s, p$_z$, and p$_x$, p$_y$) and the p$_z$ orbital of F all contribute (see Fig. \ref{FIG3}(a)). 
As in pristine graphene, the $\pi$-orbitals start below -7 eV. In Fig. \ref{FIG3}(b), we see three energy intervals with their strong peaks in the intervals (-2, -1), (-0.5, 1), (1, 2.5) eV, in which p$_z$ orbitals dominate. The outer intervals correspond to valence-$\pi$ and conduction-$\pi$* states. 
As expected, from this figure we see that the C(F)-p$_z$ orbital has no contribution to these bands. PBE-PAW predicts that these two bands have linear dispersion as in pristine graphene, which form an approximate cone at the $\Gamma$ point. At the edges of these two bands, (between -2 and 2.5 eV), orbitals with small F--(p$_x$+p$_y$) and C(F)-(p$_x$+p$_y$) character are present. Close examination of the pDOS in the middle energy window, (-0.5 to 1 eV), shows the dominant presence of p$_z$ orbitals of F, C(NN) and its NNN. The carbons are from opposite sublattices to C(F), indicating that this quasi-localized mid-gap state is induced by the presence of the F-adatom. The mid-gap band is nearly flat (about 0.71 eV bandwidth). A small contribution from overlapping C(F)-s and F-s orbitals with C(NN)-p$_z$ orbitals, lowers the energy of this band in some regions and adds to its dispersion resulting from the finite F--F distance in plane. We note that the F-induced mid-gap state has larger bandwidth relative to the H-induced state on graphene (0.06 eV)\cite{Kim}. This overlap is a result of C(F) puckering out of the plane along with the stronger electronegativity of fluorine atom. 

The total spin is quenched to roughly zero by the dilution of fractional spin distributions on all atoms, as predicted by PBE-PAW. Therefore, small spin splitting subsists in the overall band structure, and the ground state is nearly spin-unpolarized and metallic. We expect the C(F) and F-p$_z$ electrons, which are localized, to leave the system with a magnetization of 1 $\mu_B$. The strong $\pi$-conjugation in the hexagonal rings can also provide additional stability for the unpaired electron derived from fluorine adatom. In short, the M$\approx$0 result from PBE is unphysical.

Experimental findings indicate spin 1/2 paramagnetism in the partially fluorinated graphene sheet, resulting from unpaired spins in $\pi$-bands \cite{Nair_2012, Cheng}. The number of measured paramagnetic centers reported in Ref.~[\onlinecite{Nair_2012}] was not equal to the number of F adatoms in the sample, but three orders of magnitude smaller. In Ref.~[ \onlinecite{Nair_2012}], this was explained by fluorine's significant tendency to cluster on graphene, and that the bipartite nature of the graphene lattice destroys the magnetic moment forming in the interior part of the fluorine cluster. Hence, the magnetic moment contribution comes only from adatoms near the cluster edges, or only those fluorine atoms that have no counterparts on the neighboring sites of the bipartite sublattice. For an single F atom, one expects a total magnetic moment of 1~$\mu_B$ to emerge for structure like Fig.~\ref{FIG1}. 

The prediction of a metallic band structure is another conflict between PBE and experiments. All experiments confirm that FG is a semiconductor. The value of the experimental band gap has sparked a debate in the literature, and shows large discrepancies between 68 or 80 meV up to 3eV\cite{Nair_2010, Zbo, Zhao, Cheng}.  The measurements will be affected by states inside the gap due to: the number of graphene layers, impurities, defects and vacancies appear during experimental synthesis and the annealing of the structure. In addition, different fluorination approaches and relevant fluorination parameters result in different conformations and C/F ratio\cite{Cheng_2016}, which affects the electronic and physical properties of the structure. Nair et al. in particular reported fully fluorinated single-side graphene is a semiconductor with a wide gap of 3.0 eV and resistivity higher than 10$^{12}~\Omega$/m at room temperature\cite{Nair_2010}. 

We also expect the FG paramagnetic transition to be accompanied by a metal to insulator transition, as shown by experiment. While PBE fails in our expectations, one should recall that correlation effects are central to many metal-insulator transitions. Also, as discussed above, electrons are moving in a narrow mid-gap band, which implies strong electron-electron correlation. Importantly, the C(F) and F atom partial DOS in the vicinity of the Fermi level show contributions of d orbitals in this region. This behavior has also been reported in Ref.~[\onlinecite{Irmer}]. Therefore, we improve our local electronic treatment by using PBE+U to see the effects of correlation, induced by C--F bonding, in the total magnetic moment and electronic properties of FG. Moreover, by using this approach one can reduce the SIE associated to delocalization errors in DFT. The results are discussed in the next section.


\begin{figure*}
\hspace*{-0.5 cm}
\centering
\includegraphics[width=1 \linewidth]{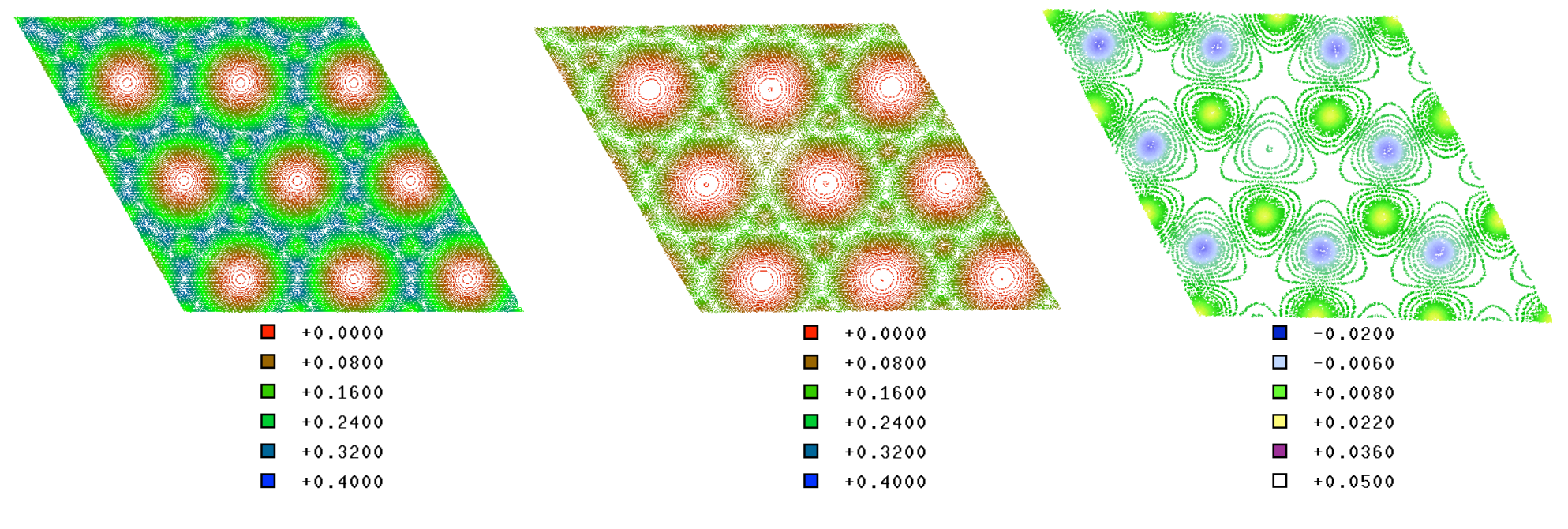}
\caption{\label{FIG4} (Color on line) (a) Contour map of the density distribution of electrons on the plane of pristine graphene shows a uniform distribution of $\pi$-electron between carbon bonds. (b) In contrast to pristine graphene, p$_z$-electron density perpendicular to the plane in FG is not distributed homogeneously. An obvious charge depletion around C(F) atom is seen. Charges are semi-localized around carbons on B sublattice, when F bonds to carbon on sublattice A (c) breaking of $\pi$-bonding between C(F)-C(NN) leads to the induction of magnetic moments throughout the lattice. The maximum positive polarization contributions are from carbons belonging to the sublattice B (denser green lines). As shown, the directions of the magnetic moments also in two sublattices are opposite, so that negative polarization contributions are from carbons belonging to the sublattice A (blue lines). 
The associated thermometer is shown below each figure.}
\end{figure*}
\subsection{PBE+U Electronic structure }\label{subsec: pbe+u}
It is well known that LDA and GGA functionals produce total energies with unphysical curvatures for fractional numbers of electrons\cite{Mori, Perdew}. This is related to the SIE stemming from artificial delocalization in Kohn-Sham occupied orbitals, and gives a spurious contribution to the total energy, hence one cannot retrieve the correlation correctly. In our study, with an open-shell atom (F) and an odd number of electrons in the metallic graphene surface, SIE produces a severe problem. The self-interaction corrected hybrid functionals (such as Heyd-Scuseria-Ernzerhof (HSE)\cite{Krukau}) can reproduce qualitatively the magnetic properties of fluorinated graphene. In Ref.~[\onlinecite{Kim}] it is shown that HSE recovers 97\% of the experimental magnetic moment, and a gap of about 0.51 eV at the  $\Gamma$ point of a single F-adatom on graphene for a $3\sqrt3\times3\sqrt3R30^{\circ}$ supercell (1:54 of fluorine coverage). The non integer M and small gap imply that this is a problematic phase with HSE too, and further tuning of the fraction of Fock exchange may be needed.

In contrast to the HF method, in DFT+U the effective interaction is orbital-specific
 and computationally much more efficient. If the U value compensates the curvature of the DFT total energy profile, DFT+U can also cure the problematic SIE\cite{Cococcioni}. 
The on-site electron-electron interactions, which here are important for localized electrons in our narrow mid-gap state, are taken from the atomic Hartree-Fock Slater integrals. The hopping of electrons between neighboring atoms is costly, giving added correlation between electrons sharing the same orbital. The on-site repulsion energy in the p-orbitals increases the cost of double occupation, the spin-degenerate bands will split near the Fermi energy, and the correct magnetization and optical gap will be retrieved. 

An important question is which atoms the Hubbard U should be applied on: the p orbitals are common to all of the atoms. We begin by applying the U correction only to the fluorine, C(F) and its three NN carbons. In this case we obtain a magnetic moment close to 1~$\mu_B$, but the calculated band structure is similar to that for PBE discussed in the previous section, with nearly zero band gap. As mentioned above, the pDOS calculations confirm that p$_z$ orbitals of most of the atoms in the supercell contribute to the mid-gap state. As a result, we apply U to electrons in the p orbitals of all carbons and fluorine in the supercell.

To determine the parameters U and J, we scan a large range of values, for their optimum with respect to experiment. We start by neglecting the lesser term in the total energy, setting $J=0$. This term originates from non-sphericity of the electronic interaction, yielding the simple  $"$+\textit{U} $"$ approach. The optimum U energy should restore the system to the expected magnetic moment near to M=1~$\mu_B$. We have performed a systematic investigation and increased U in progressive steps to find the smallest physical value. 
By setting $U=5.3$ eV the magnetic moment of M=1~$\mu_B$ is recovered (we also examine larger values). Next, by increasing J and changing U, we further searched the correlated electron space, for a total moment of M=1~$\mu_B$ obtained from the smallest U and minimal total energy. We scanned J starting with small value of 0.1 eV, and changed U in the vicinity of 5.3 eV. The adjustment has been done with special care in small energy intervals. Finally, we end up with $U=5$~eV and $J=0.1$~eV, which yield a total magnetic moment of M=1~$\mu_B$. In this condition, a direct gap of 1.08~eV opens at $\Gamma$ point between the highest occupied energy (minority electrons at -3.73~eV, or -0.23~eV wrt $E_F$) and the lowest unoccupied energy (majority electrons at -2.65~eV, or 0.85 eV wrt $E_F$). As shown in Fig.~\ref{FIG2}(b). FG is found to be an indirect semiconductor with a gap of 0.41~eV between the $\Gamma$ and M points in the minority electron channel. 
The same direct gap (1.07 eV) and M=0.999~$\mu_B$ can be obtained from the ABINIT code, using a different implementation which applies the Hubbard U in atomic spheres instead of all space (important for extended p orbitals). The parameters must be pushed to $U=12.5$~eV and $J=1.2$~eV to yield the same gap and magnetic moment (very similar overall band structure). In this case at least we can obtain the same physical results as from the DFT+U implemented in QE in all space, but with much larger U and J values, as they are applied only on a small part of the atomic-like orbitals. We note in passing that (1) the self-consistent loop shows faster and more stable convergence in the localized U implementation in ABINIT; and (2) we do not claim that this equivalence holds in general for DFT+U on extended p orbitals, as it will depend on bonding environments. 

In terms of the total energy, PBE+U shows a spin-polarized ground state about 169~meV below the spin-unpolarized state (compare with 1~meV for PBE). The band structure shown in Fig.~\ref{FIG2}(b) displays a clear exchange splitting for all bands. The maximum splitting occurs for the mid-gap state with 1.58~eV at point~\textbf{K}. The valence and conduction bands split by at maximum 0.32 eV (near point \textbf{K}) and 0.34eV (at point \textbf{K}), respectively. From Fig.~\ref{FIG2}(b), we observe that the characteristic linear dispersion of graphene has been modified. The split mid-gap state explains elegantly the magnetic state, with the occupied spin-up mid-gap state degenerate with the valence band at $\Gamma$, and the unoccupied spin-down state degenerate with the conduction band. This is consistent with the experimental finding that even weakly fluorinated graphene is an insulator with three orders of magnitude higher resistivity than graphene at room temperature \cite{Nair_2010}.  
In spite of a different fluorination ratio, our DFT+U electrical features shown in Fig.~\ref{FIG1} for a 3$\times$3 supercell is comparable to the HSE hybrid functional band structure for a fluorine adatom in a $3\sqrt3\times3\sqrt3R30^\circ$ graphene supercell from Ref.~[{\onlinecite{Kim}]. A direct band gap of 0.51 eV at point $\Gamma$ and total magnetization of 0.97~$\mu_B$ is predicted for this supercell. Their structure is also predicted to be an indirect gap semiconductor. 

 The GGA+U predicted pDOS of p$_z$ orbitals for four different carbon atoms, on both A and B sites, are shown in Fig.~\ref{FIG_SI} of supplementary information. By convention we will call the spin on F ''up''. PDOS of p$_x$ and p$_y$ orbitals show no polarization from both PBE and PBE+U and are not shown here. In GGA, almost no spin splitting for p$_z$ orbitals is observed, and no net magnetization is predicted.  Instead, in GGA+U, there is a clear spin splitting of the p$_z$ orbitals, and there is a net spin up. The difference between the pDOS for the two sublattices is another distinctive feature can be observed in Fig.~\ref{FIG_SI} of supplementary information. As shown in Fig.~\ref{FIG1}, atoms C4 and C10 belong to the same sublattice A as C(F), while atoms C9 and C14 belong to sublattice B. 
There remains a small down spin for atoms on sites A, which does not cancel the resultant up polarization from sites B, leading to a ferrimagnetic state with net spin of 1~$\mu_B$ for FG.

\begin{figure}
\hspace{-0.3 cm}
\centering
\includegraphics[width=1. \linewidth]{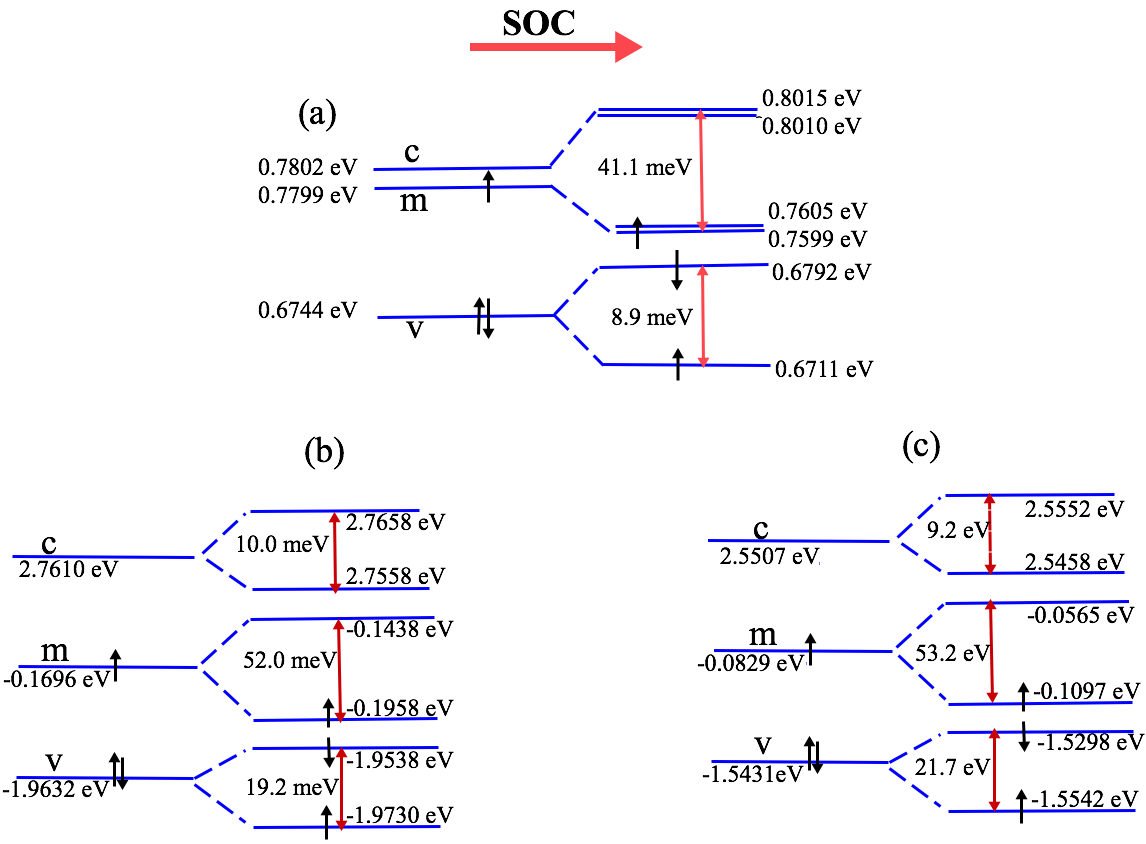}
\caption{\label{FIG5} {(Color on line) Energy levels and SOC splitting predicted by “spin-unpolarized” PBE+U and PBE+U+SOC calculations for (a) point $\Gamma$ , (b) point \textbf{M}, and (c) point\textbf{ K}. The maximum splitting of mid-gap state is at $\Gamma$. The Fermi energy of the spin-unpolarized PBE+U structure is set to zero.\\
}}
\end{figure}

\begin{figure}
\hspace{-0.4cm}
\centering
\includegraphics[width=1.03 \linewidth]{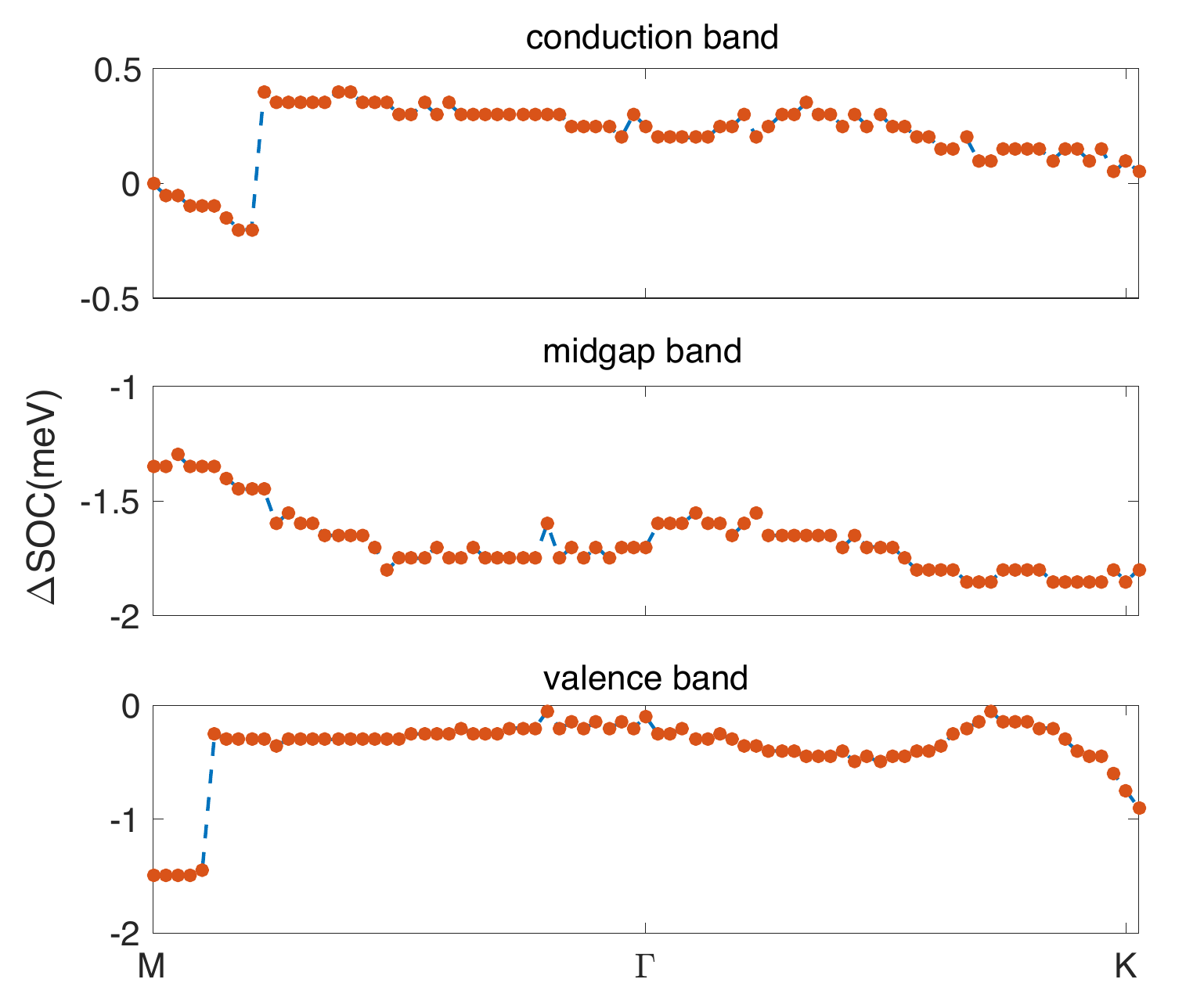}
\caption{\label{FIG6} {(Color on line) SOC splitting energies of (a) valance, (b) mid-gap and (c) conduction bands predicted by non-collinear “spin-polarized” PBE+U+SOC calculations. The jump near M is due to a band crossing.}}
\end{figure}

 GGA+U predicted pDOS of C(F) and F are also shown in Fig.~\ref{FIG_SI}. Both C(F) and F atoms have net spin up, however as the p$_z$ electron of F and C(F) are involved in $sp^3$ bonding, the polarization of these atoms are small (0.0304 and 0.0748 $\mu_B$ for C(F) and F, respectively). The C(F) NN have the maximum positive polarization of 0.25 $\mu_B$. The polarization decreases slowly with distance in this sublattice (0.23 $\mu_B$, 0.15 $\mu_B$ for C9 and C13, respectively). Comparing the contour maps of charge density distribution of pristine graphene and FG shows two different patterns: As it is seen in Fig. \ref{FIG4}(a), the charge density is distributed uniformly between carbons bonds in pristine graphene. While, in Fig. \ref{FIG4}(b), the introduction of a fluorine atom on sublattice A breaks the aromatic bonds in sublattice B around the tetrahedral C(F) atom. In addition, electron depletion around the C(F) atoms makes the charge distribution different in the two sublattices. Consequently, $\pi$-orbitals are not completely delocalized, and electrons do not propagate freely over hexagonal rings. As U increases, these localizations are followed by spin ordering and a metal-insulator phase transition, as seen in Ref.~[\onlinecite{Hong}]. The contour map of magnetization distribution in the graphene plane obtained from PBE+U is shown in Fig.~\ref{FIG4}(c), with a clear antiferromagnetic radial oscillation around the F atom. The maximum on the nearest atoms to the central carbon, on sublattice B, is followed by a maximum down spin ($\sim$-0.15~$\mu_B$) on sublattice A. 
 
 \subsection{Spin-orbit splitting and magnetic anisotropy}
Finally, we have performed non-collinear PBE+U+SOC calculations with fully relativistic pseudopotentials with initial magnetizations which were either (1) zero (to see pure SOC splitting) and (2) non-zero (to test the ground state collinear structure).
 
(1) Spin compensated PBE+U predicts FG to be metal, with a degeneracy between the mid-gap state and conduction band at $\Gamma$, as shown in Fig.~\ref{FIG5}.  Incorporating SOC lowers the symmetry and the energies of these two bands. Here, the SOC-assisted gap is 41.1 meV. Likewise, large SOC exchange splittings are predicted for the mid-gap state at the \textbf{K} and \textbf{M} points by 53.2~meV and 52.0~meV, respectively. 

(2) PBE+U clearly demonstrates nearest-neighbor spin interactions, originating from the presence of the F-adatom. Here, $\Delta E_{SOC}$  is calculated as the additional shift between a pair of bands after their initial spin splitting. The results are shown in Fig.~\ref{FIG6}. Our ''spin-polarized'' non-collinear PBE+U+SOC calculations yield only small differences compared to the collinear valence, mid-gap and conduction bands presented in Fig.~\ref{FIG2}(b). With SOC the total magnetic moment reduces only about by 2\% from its initial value and a small canting is found away from the normal direction (0.16~$\mu_B$ in plane). The maximum $\Delta E_{SOC}$ splitting of about 2~meV occurs at valence and mid-gap states. The FG structure shown in Fig.~\ref{FIG1} has no inversion symmetry and SOC splitting appears throughout the BZ, while $\Delta E_{SOC}$ is approximately constant through different \textbf{k} points, as shown in Fig.~\ref{FIG6}. Our PBE+U+SOC patterns are different from those presented by Ref.~[\onlinecite{Irmer}], in which there is no SOC splitting at $\Gamma$. With DFT+U, since the exchange interaction splits majority and minority electrons at $\Gamma$ point, (E(k=0,$\uparrow$)$\neq$ E(k=0,$\downarrow$)), therefore SOC splitting appears here too as seen in Fig.~\ref{FIG6}.

The SOC effect in pure graphene layers is very weak (about 25~$\mu$eV\cite{Kane}). It is well known that lattice distortion induced by adatoms giving rise to a much larger SOC within graphene layers\cite{Neto}: the SOC approximately depends on the amount of sp$^3$ hybridization of the carbon atom at the puckered out point. Following the theory introduced in Ref.~[\onlinecite{Neto}], we estimate relative strength of the SOC coupling with:
 \begin{equation}
\frac{\Delta E_{SOC}(\theta)}{\Delta E_{SOC}(at)}=\frac{\sqrt{2}}{\tan^2\theta }\sqrt{6(\tan^2\theta-2)},
\end{equation}
where $\Delta E_{SOC}(at)$ shows carbon atomic SOC ($\sim$10 meV), and $\theta$ denotes for the angle between the new $\sigma$ bond of FG structure and the normal direction along C--F bond. The PBE predicted induced distortion angle of 102.46$^{\circ}$, should increase $\Delta E_{SOC}$ approximately to 0.51\%  of the atomic value, ie to about 5 meV. This estimation is of the same order of the 2~meV calculated by PBE+U. 

We inspected the band structures for possible Rashba features along different \textbf{k}-paths in the BZ, but find no important shifts.
The magnetic properties are set by the spin-exchange energy, $\Delta E_{ex}>>\Delta E_{SOC}$ , induced by fluorine adatoms. 
 
 A SOC splitting value of the same order of magnitude as what we find from PBE+U (5.1~meV and 9.1~meV) has been measured under $\sim$0.005\% and $\sim$0.06\% fluorination \cite{Avsar}, while a previous DFT effort predicted much larger splitting of a few tens of meV~\cite{Irmer}. 

Finally, we calculated the MAE, which reflects the orientation dependence of the total energy of the ground state of the system arising from SOC. MAE is calculated from the change in total energy with respect to variation of magnetic moment direction from the in-plane (x) into the perpendicular (z) directions. In most bulk ferromagnetic materials, magnetic anisotropy in a particular direction is a consequence of spin-orbit coupling. The perpendicular direction (along the C--F bond) is found to be the preferred orientation by 3.95~$\mu$eV. The MAE is expected to increase at lower fluorine concentration, as observed experimentally \cite{Hong}.\\

\section{Conclusion}
In summary, we show that PBE cannot correctly determine the magnetic and electronic features of fluorine adatoms on graphene. Instead, we have demonstrated that PBE+U provides a good description of this system, and even presents specific advantages over hybrid functionals. We obtain optimum U and J parameters values of 5 and 0.1~eV in the GGA+U calculation. We find that fluorine adatoms on graphene produce localized p$_z$ electrons throughout the lattice, and the system becomes ferrimagnetic and strongly correlated. The electronic and magnetic properties presented in this work are dominated by the influence of F-adatoms, but spread to both sublattices in the graphene substrate. The magnetic moment induced by fluorine plays a central role to control electronic and magnetic properties, and SOC has a minor effect on the electronic band structure. A magnetic anisotropy energy of 3.95~$\mu$eV is found with an out of plane easy axis. Our results show exchange interaction induced by fluorine is dominant, and no Rashba shift is found. 
\section{ACKNOWLEDGEMENT}

MJV acknowledges funding by the Belgian Fonds National de la Recherche Scientifique FNRS under grant number PDR T.1077.15-1/7, and support from ULg and from the Communaut\'{e} Fran\c{c}aise de Belgique (ARC AIMED 15/19-09).
\section{APPENDIX: comparing PBE+U with PBE pDOS of atoms }
 As discussed in Sec.~\ref{subsec: pbe+u}, pDOS of atoms predicted by PBE functional does not show exchange splitting, either belonging to sublattice A or B. In contrast, by including correlation term between electrons, PBE+U shows clear exchange splitting for atoms especially if belonging to sublattice B, as seen in Fig.~\ref{FIG_SI}.
 \begin{figure*}
\centering
\includegraphics[width=0.8 \linewidth]{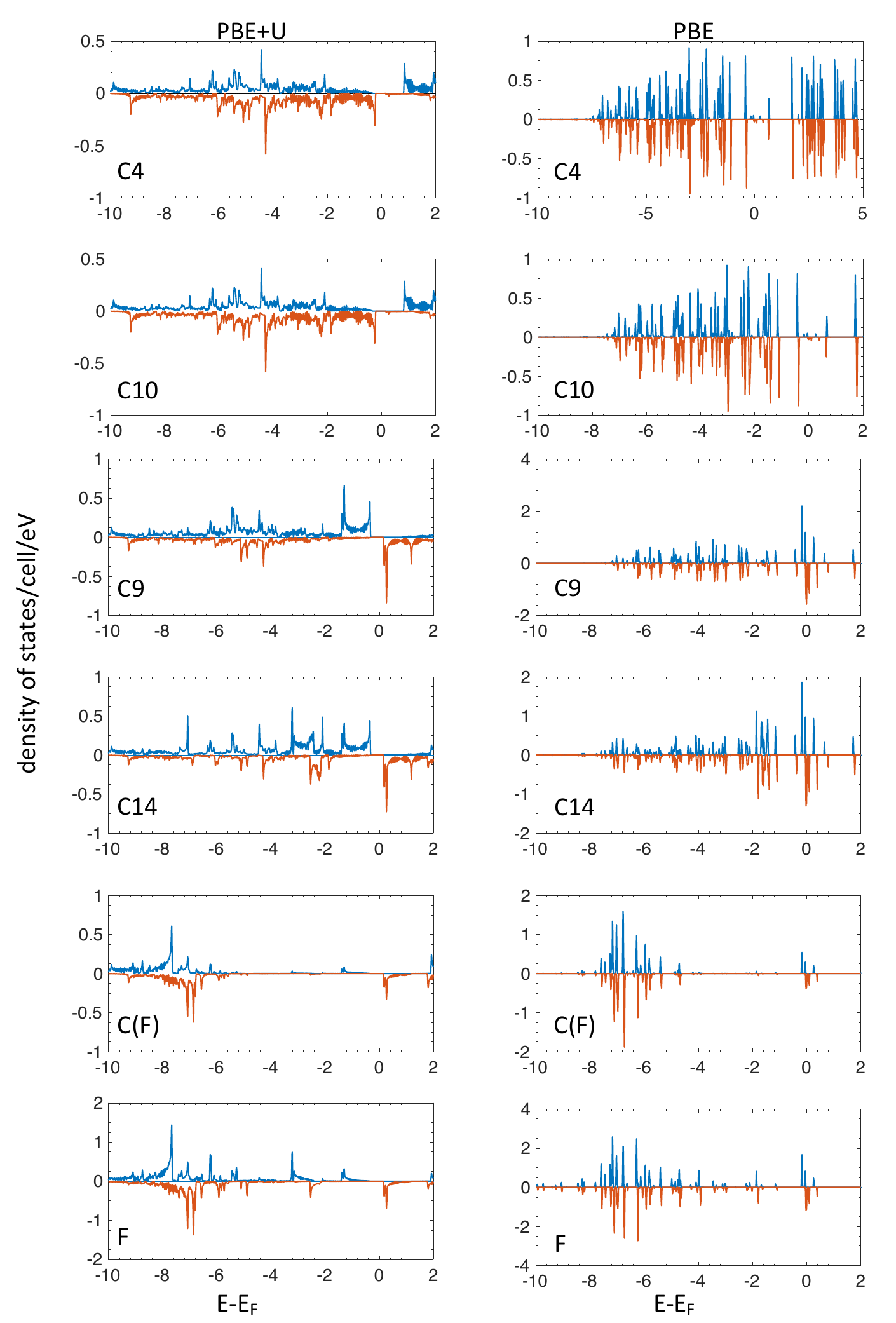}
\caption{\label{FIG_SI} (Color on line) PBE+U (left columns) and PBE (right column) predicted pDOS of majority (blue color) and minority (red color) electrons of p$_z$ orbitals corresponding to different carbon atoms as typical examples. C4, C10  and C(F) in which bonded to fluorine (F) atom are in sublattice A, C9 and C14 in sublattice B of the supercell as shown in Fig.~\ref{FIG1} of the paper. Comparing PBE+U pDOS of occupied p$_z$ and p$_x$ orbitals of majority and minority electrons shows the magnetic properties of FG mostly stemming from p$_z$ orbitals of atoms in sublattice B. Moreover, small negative plarizations from p$_z$ orbitals in sublattice A are observed. Almost no plarization for p$_z$ orbitals is resulted from PBE calculations.  Fermi energies E$_F$ are set to zero.}
\end{figure*}
\newpage


\end{document}